\documentstyle[12pt,epsf,epsfig]{article}
\newcommand{\NN}{
\begin{picture}(11,3)(-2,-2)
\put(1,-12){$\tilde{}$}
\put(-2,-2){$N$}
\end{picture}}

\begin{document}
\begin{center}
%\vspace*{1cm}
\LARGE
A proposal for analyzing the classical  \\
limit of kinematic loop gravity.\\

\vspace*{1.5cm}

\large

Madhavan Varadarajan$^{*}$
\\and\\Jos\'e A. Zapata$^{*,**}$.
\vspace*{1.5cm}

\normalsize
$^*$Raman Research Institute\\
Sir C. V. Raman Avenue\\
Bangalore 560 080, India\\\hspace{5mm}\\
$^{**}$ Instituto de Matam\'aticas, UNAM\\
A. P. 61-3, Morelia, Mich. 58089, Mexico\\

\vspace{.3in}
May24, 2000\\
\vspace{.3in}
ABSTRACT
\end{center}
We analyze the classical limit of kinematic loop quantum gravity in
which
the 
diffeomorphism and hamiltonian constraints are ignored. We show that
there 
are no quantum states in which the primary variables of the loop
approach,
 namely the SU(2) holonomies along {\em all} possible loops,
approximate their classical counterparts. 
At most a countable number of loops must be specified.
To preserve spatial covariance, we choose this set of loops to be
based on  physical lattices specified by the quasi-classical 
states themselves. 
We construct ``macroscopic'' operators based on such lattices and propose 
that these operators be used to analyze the classical limit. 
Thus, our aim is to
approximate classical data using states in which appropriate
macroscopic 
operators have low quantum fluctuations.

Although, in principle, the holonomies of `large' loops on these lattices 
could  be used to analyze the classical limit, we argue that it may be 
simpler to base the analysis on an alternate set of ``flux'' based 
operators.
 We explicitly construct candidate 
 quasi-classical states in 2 spatial dimensions and indicate how these 
constructions may generalize to 3d.  We discuss the less robust aspects of
our proposal with a view towards possible modifications.
Finally, we show that our proposal also applies to the 
diffeomorphism invariant Rovelli model which couples a matter reference 
system to the Hussain Kucha{\v r} model.

\vspace*{1cm}
%\noindent PACS number(s): 04.20.Cv, 04.20.Fy

\pagebreak

\setcounter{page}{1}        

\section*{1. Introduction}
The loop quantum gravity approach has yielded a number of interesting
results. A mathematical arena 
has been defined in which the constraints of quantum gravity have been
expressed as quantum operators. The complete  kernel of the 
diffeomorphism constraints has been obtained \cite{mafia} and efforts
are
on to find the kernel of the Hamiltonian constraint
\cite{thomas,ReiRov}
However, contact with 
the classical limit (i.e. general relativity) has been elusive.

Since very little is known regarding the interpretation of the kernel
of the constraint operators,
the unambiguous results pertaining to the classical limit 
have been obtained at the kinematic level wherein 
the diffeomorphism and Hamiltonian constraints are ignored
\cite{areaRS,areaAL,weave}.
By taking recourse to the arguments of Rovelli
\cite{carlomatter,carloHK}, 
it is,
however, not inconceivable that kinematic results may be physically
relevant.
Moreover, in any situation with classical boundary conditions 
(e.g. black hole horizons, asymptotically flat spacetimes),
%
%\footnote{In the context of asymptotically flat spacetimes 
%the framework constructed in \cite{sameer} may be relevant. 
%}), 
the classical 
constraint vector fields  leave the boundary conditions invariant.
Hence,
at 
the boundary, the smearing functions (lapse and shift) for the
constraints
typically vanish and kinematic results may acquire physical
significance. 

Even at the kinematic level, almost all work to date is restricted to an 
exploration of the classical limit of (functionals of) 
the spatial metric or densitized triad operators
\cite{weave,carlojunichi}.
\footnote{An exception is 
\cite{arnsdorf}. 
} 
 `Weave' states have been constructed which 
approximate classical metrical information. 
It is possible that the conjugate (connection) variable fluctuates
wildly 
in such states and, if so, these states cannot be quasi-classical.

In this work we propose a framework to analyze both the metrical as well as the connection  degrees of 
freedom with a view towards the classical limit.
The reason a new framework is required is as follows.

The  connection dependent operators which have unambiguous classical 
counterparts are the traces of holonomies around loops. The latter are
denoted by 
$T^0_{\gamma}(A) $ with 
\begin{equation}
T^0_{\gamma}(A) = {1\over 2}Tr H_{\gamma} (A), \;\;\; 
H_{\gamma} (A) = P\exp - \oint_{\gamma} A_a dx^a,
\label{holdef}
\end{equation}
where 
$\gamma$ is a loop embedded in the spatial manifold $\Sigma$,
$A_a$ is an SU(2) connection,  $H(\gamma )$ is the holonomy and $Tr$
denotes
the trace in the $j={1\over 2}$ representation.
It would be natural to explore the classical limit in terms of these
operators. Then, quasiclassical states would be required to approximate
the set of all holonomies and say, surface areas, through quantum expectation
values of the corresponding operators with low fluctuations. Unfortunately,
as we show in the beginning of section 2, holonomies of a classical connection
along all possible loops cannot be approximated by any quantum state!
More precisely, it is only on a countable set of loops (the set of all loops 
is, of course, uncountable) that holonomies have a chance of being 
approximated. However, an arbitrary choice of this countable set is in 
conflict with spatial covariance. 

Therefore, a new framework is needed to analyze the connection degrees of 
freedom. In this work we propose such a framework. Our main ideas are 
as follows. Any quasiclassical state must approximate, in some way, the 
data corresponding to both the spatial metric as well as the connection.
To approximate a given spatial metric we need states 
defined on a ``large enough'' graph. 
We require that this  graph  gives a  
latticization of the (compact) spatial manifold.  Then the preferred
set of loops are naturally identified as those which lie on the lattice.
We make these ideas precise in section 2 in such a way that the resulting
framework is spatially covariant.

Next, given the set of loops on a lattice, we would like to approximate
a classical connection. The natural set of operators to consider are the 
holonomies along these loops. Since we  are interested in approximating 
classical behaviour at scales much larger than the Planck length,
 it is enough to restrict attention to  
loops of size much larger than the Planck scale (the size of a loop is 
measured by the metric part of the classical data).  
Thus one natural set of
connection operators for an analysis of the classical limit are the 
holonomies along large  loops which lie on the lattice. However, the 
lattice structure suggests an alternative set of operators. These are the 
`magnetic flux' operators of lattice gauge theory which measure the 
non-abelian magnetic flux through the plaquettes of the lattice. They are 
constructed in the usual way from holonomies along the plaquettes.
For reasons which we spell out in section 3, we choose to
base our analysis of the classical limit on these operators rather than the 
large loop holonomies. We devote section 3 to this change of focus
from holonomy operators to flux operators.

In section 4 
we work out our ideas in detail for the case of two spatial 
dimensions and explicitly display 
states 
which approximate aspects of both 
the classical spatial metric and the $SU(2)$ connection.
We also indicate how our constructions can be extended to the case of 
three spatial dimensions.

Section 5 is devoted to a discussion of various issues
which arise in the context of our proposal, with an emphasis on its less 
robust aspects. The discussion in this section 
indicates that some of our ideas are too simplistic whereas
others posess attractive features; it thus points to ways in which the
proposal may be modified.
We also show, in section 5, how our proposal can be extended to the
diffeomorphism 
invariant context of Rovelli's work \cite{carloHK} wherein the 
Hussain Kucha{\v r} model is coupled to a matter reference system.
Section 6 contains our conclusions.

There seems to be no single viewpoint with regard to the role, 
within the framework of loop quantum gravity, of considerations at the purely
kinematic level.
 Therefore it is appropriate that we spell out our viewpoint before describing
our results.

The aim of loop quantum gravity is to construct a quantum theory which
has general relativity as its classical limit. Since, in this approach,
 the Hamiltonian 
constraint operator is poorly understood, it would be premature to discuss the
classical limit at the full dynamical level. Even at the (spatial) 
diffeomorphism invariant level, with the exception of the 
total volume operator, 
quantum operators corresponding to 
diffeomorphism invariant classical observables have not been constructed.
Without these operators it is
difficult to interpret the theory and discuss its classical limit.
Since even the kinematic state space is very different from 
that of conventional flat space quantum field theory, it makes sense to
understand the classical limit first at this kinematic level, where even the 
diffeomorphism constraints are ignored.
The classical limit consists of smooth metric and connection data. The 
approximation of smooth metrical data by weave states is already subtle and
the approximation of smooth connection data is still an open question.
It is our view that an analysis of the classical limit at the kinematic level
may clarify strategies for analysing the classical limit at the spatial
diffeomorphism invariant level and finally, (once the Hamiltonian constraint 
is well understood) at the fully dynamical level.

Independent of the above `structural' role of understanding the classical 
limit of kinematic gravity, is the question of whether results at the 
kinematic level have any relevance 
to {\em physical} predictions of full blown 
quantum gravity. 
\footnote{Note that  we are not considering those special situations
mentioned earlier in this section, involving boundary conditions,  
where kinematic results are already `gauge invariant'.}
 For example, the 
discreteness of the spectrum of the area operator is often cited by some 
workers as a physical prediction.
Since the Hamiltonian constraint is not well understood, we refrain from
discussing this issue in the context of full quantum gravity.
Instead, we restrict our attention to the possibility of promoting kinematic 
results to predictions at the diffeomorphism invariant level.
%A diffeomorphism invariant state is obtained by diffeomorphism averaging a 
%given spin network state \cite{mafia}. A seemingly plausible strategy  
%to make contact between kinematic results and diffeomorphism invariance, is to
%choose a diffeomorphism invariant state based on a knot class and then
%`gauge fix' so that a kinematic state based on a particular graph of this
%knot class represents the chosen diffeomorphism invariant state. We do not 
%subscribe to this strategy for the following reason. The diffeomorphism
%invariant state space is obtained through Dirac quantization as opposed to
%gauge fixing. Gauge fixing is a notion appropriate to classical theory - it 
%cannot be done at the quantum level 
%after imposing quantum constraints through 
%the Dirac procedure. Gauge fixing of the diffeomorphism constraint at the
%classical level {\em cannot} lead to the kinematic Hilbert space of loop 
%quantum gravity because the very construction of the latter is predicated
%on considerations of diffeomorphism invariance of the Hilbert space measure, 
%as opposed to breaking 
%diffeomorphism invariance via gauge fixing.
One way to promote kinematic results to the 
diffeomorphism invariant level, is to couple the gravitational variables to
a matter reference system as in, for example, \cite{carloHK}.
This can be done only if the kinematic framework for the gravitational 
variables is spatially covariant. In what follows, we shall be guided by this
requirement of spatial covariance.

\noindent{\em Notation and Conventions}: We assume familiarity with the
loop 
quantum gravity approach (for example see \cite{mafia} and references 
therein) and use notation which is standard in the field. $a,b..$ are 
spatial indices, $i,j..$ are internal $SU(2)$ indices, $A_a^i(x)$ is
the 
$SU(2)$ connection and ${\tilde E}^a_i(x)$ is the densitized triad.\\
\noindent
$\{A_a^i(x),{\tilde E}^b_i(y)\} = \iota G_0\delta_a^b \delta (x,y)$
where 
$\iota$ is the (real) Immirzi parameter
\cite{immirziPar}. We shall restrict attention
to
piecewise analytic loops/graphs.

$\overline{\cal A}$ is the completion (via a projective limit 
construction) of the space of smooth connections ${\cal A}$,
$\overline{{\cal A}/{\cal G}}$ is the
Gel'fand completion of the space of smooth connections modulo gauge and
$d\mu_H$ denotes the Ashtekar-Lewandowski (or Haar) measure on
$\overline{\cal A}$ as well as on $\overline{{\cal A}/{\cal G}}$. 

$\hat O$ is the operator version of the classical object $O$, 
${\hat O}^{\dagger}$ is its adjoint and $O^*$ is the complex conjugate 
of $O$. We shall often denote the expectation value of 
 $\hat O$ in the quantum state under discussion as 
$<{\hat O} >$.
$l_{0P}$ is the length constructed from the dimension-full gravitational 
coupling 
$G_0$, $\hbar$ and $c$. In 3+1 dimensions,
$l_{0P}=\sqrt{G_0\hbar\over c^3}$. We shall 
use units in which $\hbar = c=1$.

\section*{2. The necessity for a new framework and a sketch of our proposal}

The most straightforward approach to an analysis of the classical 
limit of loop quantum gravity would be to construct minimum
uncertainty 
states for the basic operators of the theory. These operators are the 
`configuration' operators, ${\hat T}^0_{\gamma}$, and suitable
`momentum'
operators. The latter may be 
chosen as the area operators, ${\hat A}_S$ \cite{areaRS,areaAL} 
($A_S$ is the 
area of a surface $S$ in $\Sigma$).

A tentative definition of a quasi-classical state as a minimum
uncertainty 
state for this set of operators is as follows. A kinematic
quasi-classical
state $|\psi > \in L^2(\overline{{\cal A}/{\cal G}}, d\mu_H)$ 
which approximates 
the $SU(2)$ gauge equivalence class of the classical data, 
$(A_{0a}^i(x), {\tilde E}^b_{0i}(x))$, is such that, for all 
$\gamma , S$\\
\noindent (i) $|<{\hat T}^0_{\gamma} >-
               T^0_{\gamma}(A_0)|$ and 
$\Delta {\hat T}^0_{\gamma}= 
     \big(<({\hat T}^0_{\gamma})^2 >-
        <{\hat T}^0_{\gamma}>^{2}
     \big)^{1\over 2}$ are small. \\
\noindent (ii)  $|<{\hat A}_S >- A_S(E_0)|$
and $\Delta {\hat A}_S$ are small compared to $ A_S(E_0)$.

Since $|T^0_{\gamma}(A)| <1$, we interpret `small' in (i) as 
`small compared to 1'. 

We now show that no quantum state exists in the kinematic Hilbert space
for which (i) is true for all loops $\gamma$.
The kinematical Hilbert space, 
$L^2(\overline{{\cal A}/{\cal G}}, d\mu_H)$, is spanned by the 
set of cylindrical 
functions, each of which is labelled by a piecewise analytic, closed,
finite 
graph. Hence any element of the Hilbert space is associated with at
most
a countable infinite set of closed graphs. From the properties of 
$d\mu_H$ it is easy to see that given any such state 
$|\psi > \in L^2(\overline{{\cal A}/{\cal G}}, d\mu_H)$, and any  loop
$\alpha$ which does not belong to the countable set of graphs
associated
with $|\psi > $, $<\psi |{\hat T}^0_{\alpha}|\psi >=0$ and 
$\Delta {\hat T}^0_{\alpha}= {1\over 2}$. Clearly,  there are uncountably
many loops of the type  $\alpha$. 
It follows that there
is
no
state for which (i) holds for all loops $\gamma$ in $\Sigma$.

Hence, the most straightforward approach to an analysis of the connection
degrees of freedom fails and a different approach, which relaxes (i) in 
some way, needs to be formulated. The remainder of this section is devoted to
the construction of such an approach.

>From our arguments above, it is clear that we must relax (i) to hold for
at most a countable set of loops. It is reasonable to require that the 
structure of this set of loops be such that we can use them to approximate,
in some way, {\em any} given loop.
A lattice structure is one which has this property.
Thus, we are naturally led to require that the set of loops for which
(i) is imposed provides a latticization of the compact spatial manifold.
(For simplicity, we shall restrict attention to lattices with a finite, though 
arbitrarily large, number of links).

However, an arbitrary fixed choice of such a lattice 
(or indeed, of any other countable set
of loops) introduces a preferred structure into the description and hence
 breaks spatial covariance.\footnote{Diffeomorphisms are unitarily 
represented on the kinematic Hilbert space. Spatial covariance implies that
classical data sets differing by the action of  diffeomorphisms
are approximated by quantum states which differ by the action of the
corresponding unitary operators.}
We get around this difficulty as follows.

It is essential to note  that we are only interested in quantum states
which approximate classical data. In particular such states approximate
the
data for the classical spatial metric. States which approximate {\em
only}
the spatial metric have been constructed in \cite{weave,grot} and 
are based on an underlying graph. If this graph does not extend into a
region $R \subset \Sigma$ then $R$ has zero volume and any surface in 
$R$ has zero area. Hence
a state based on such a graph  does not correspond to 
any  classical metric in the region $R$. It follows that the graph
underlying a quasi-classical  state  must `extend into all of $\Sigma$'
in order to approximate a classical metric on $\Sigma$.
Such graphs
are called weaves \cite{weave,grot}. 
For our purposes it seems natural to require that the 
graph underlying any weave state which 
not only approximates a classical 3- metric, but also approximates a 
classical connection (modulo $SU(2)$ gauge), provides a latticization
of
$\Sigma$. 
More precisely, we require that the graph be the 1-skeleton of some 
cellular complex \cite{fritsch-piccinini} whose topology is that of $\Sigma$. 
\footnote{Most of the weaves constructed in the literature 
(see \cite{grot} and references therein) are the disjoint union of
sets
of loops, and do not provide a latticization of $\Sigma$. 
Notable exceptions are the boundary data of spin foam models 
(see \cite{Rei,BC,Baez}). 
} Thus, the required lattice structure is not chosen arbitrarily 
but is obtained from the quasiclassical state itself. It is this feature which
preserves the spatial covariance of the resulting framework.

The availability of a lattice structure enables us to analyse many more 
functions than just the holonomies. More precisely, any function which admits 
a lattice approximant may be analysed using techniques from 
lattice gauge theory. Therefore, we shall develop our framework in such a way
as to deal with any function of the classical data which admits a lattice 
approximant (the degree of approximation will be made quantitative shortly).

To make these ideas more precise, we define the following mathematical 
structures.
Let $L$ denote a finite piecewise analytic graph which
provides
a latticization of the compact manifold $\Sigma$. Note that $L$ belongs
to
an uncountably infinite label set, since the action of a 
diffeomorphism on $L$ produces a lattice $L^{\prime}$ which is,
in general, different from $L$. 

We define the lattice projector
${\hat P}_L$ as the projection operator which maps any state in 
$L^2(\overline{{\cal A}/{\cal G}}, d\mu_H) \subset
L^2(\overline{{\cal A}}, d\mu_H)$ to its component in the subspace
spanned by spin network states \cite{snRS,snBaez} 
which have the following properties:
(a) every spin network state in the subspace is labelled by the 
graph $L$, and (b) for every such spin network state, 
 every link of the graph $L$ is labelled by 
some non-trivial (i.e. $j\neq 0$) representation of $SU(2)$.
It can be checked that 
\begin{equation}
{\hat P}_L{\hat P}_{L^{\prime}}= \delta_{L,L^{\prime}}{\hat P}_L
\label{eq:pl}
\end{equation}
where $\delta_{L,L^{\prime}}= 0$ if $L\neq L^{\prime}$ and 
$\delta_{L,L^{\prime}}= 1$ if $L=L^{\prime}$. Also
${\hat P}_L$ is a (bounded) self adjoint operator on
$L^2(\overline{{\cal A}/{\cal G}})$ so that 
\begin{equation}
{\hat P}_L = {\hat P}^{\dagger}_L.
\end{equation} 

Denote the space of 
  finite  linear combinations of spin networks
associated with all the graphs contained in the graph $L$ by 
${\cal D}_L$ and its completion in $L^2(\overline{{\cal A}/{\cal G}},
d\mu_H)$
as ${\cal H}_L$.%
\footnote{Note that, since 
all  spin network states
based on $L$ (including 
those with some or all links labelled by $j=0$) are  contained in
${\cal H}_L$ , 
${\cal H}_L \neq {\hat P}_L (L^2(\overline{{\cal A}/{\cal G}},
d\mu_H))$.
}
Note that ${\cal H}_L$ is the Hilbert space of $SU(2)$
lattice gauge theory on the lattice $L$. 

Let ${\hat O}_L$ be a bounded
self adjoint operator on ${\cal H}_L$ (or a densely defined symmetric
operator
on ${\cal D}_L$). Then define the operator 
${\hat O}$ as
\begin{equation}
{\hat O} := \sum_L {\hat P}_L {\hat O}_L {\hat P}_L .
\label{eq:defo}
\end{equation}
Here, the sum is over all possible latticizations of $\Sigma$.
${\hat O}$ has the following well defined action on any spin network
state
 in 
$L^2(\overline{{\cal A}/{\cal G}}, d\mu_H)$. Every spin network state
is associated with some  unique `coarsest' graph i.e. the graph which
has all its edges labelled by non zero spin. 
Let $\gamma_0$ be the coarsest graph for the spin network state
$\psi_{\gamma_0}$. Then, if $\gamma_0$ does not provide a latticization
 of $\Sigma$, from (\ref{eq:defo}), ${\hat O}\psi_{\gamma_0}=0$
otherwise
${\hat O}\psi_{\gamma_0}= {\hat P}_{\gamma_0}{\hat O}_{\gamma_0}
\psi_{\gamma_0}$. This action can be extended by linearity to the dense
set of finite linear combinations of spin network states in 
$L^2(\overline{{\cal A}/{\cal G}}, d\mu_H)$ and thus 
${\hat O}$ is 
a densely defined operator on this dense domain.

We now use (\ref{eq:defo}) to encode our ideas  
for the approximation of classical data $(A_{0a}^i(x), {\tilde
E}^b_{0i}(x))$.
Let the classical metric constructed from  ${\tilde E}^a_{0i}(x)$ 
be $q_{oab}$. Let $O_L$ be the classical lattice approximant to the
(real)
classical quantity $O$ on the lattice $L$. 
Typically, for classical functions of interest, the  lattice function 
$O_L$ is a sum over the `cell' functions $O_{I_L}$ where $I_L$ labels
the 
cells/plaquettes of the lattice $L$.
 The finer  
the lattice $L$,  the closer is $O_L$ to the continuum function $O$ and
the
larger is the number of `cell' contributions to $O_L$. 
The degree to which $O_L$ approximates $O$ 
can be made quantitative in terms of 
the length of the lattice parameters of $L$ as measured by $q_{oab}$.

Let ${\hat O}_L$ be the operator corresponding to $O_L$. We require
that  
${\hat O}_L$ be constructed as a self adjoint 
operator on ${\cal H}_L$ (or ${\cal D}_L$),
 from  magnetic flux type operators of 
$SU(2)$ lattice gauge theory on the lattice $L$.
Then, {\em for calculations of expectation values in a quasi-classical
state} 
we interpret 
(\ref{eq:defo}) as the operator corresponding to the classical quantity
$O$. 
Recall that we require quasi-classical states to be associated with 
some lattice $L$. From the considerations of \cite{weave}, it is 
expected that the typical link
size of such a lattice as measured by $q_{oab}$
 is of the order of the Planck length. Thus, the only term to
contribute 
to an expectation value in a quasi-classical state
in the right hand side of  (\ref{eq:defo}),  will be one associated
with
a lattice with Planck size lattice parameters! 

This completes the description of our proposed framework but for one 
last issue.
Since the operator ${\hat O}$ has the lattice projection operators,
$P_L$,
in its definition, it is not obvious that the usual 
 correspondence is guaranteed between the Poisson brackets of
macroscopic
classical quantities $O$ and the commutators of the corresponding 
operators $\hat O$. Thus it must be checked if this correspondence
holds in 
expectation value in order that  our candidate quasi-classical states be
physically acceptable.

We can now summarize our proposed framework for analysing quasiclassicality 
as follows:\\
\noindent (1) We require that any quasi-classical state
$\psi$, which approximates both 
the classical 3- metric  and the conjugate connection, $(A_0,E_0)$, 
be associated with 
some  lattice $L_0$, so that ${\hat P}_{L_0} \psi = \psi$.
\footnote{Though we shall not do so here, it seems natural to relax this 
condition and only require (2) and (3) of any quasiclassical state.}  \\
\noindent (2) Given a classical function, $O$, 
we construct a corresponding
operator ${\hat O}$ as follows. We identify the lattice approximant
$O_L$
to $O$ and construct the operator ${\hat O}_L$ in the lattice gauge
theory
on $L$. Then we construct $\hat O$ as in (\ref{eq:defo}).
\\
\noindent (3) We require that $<O>\sim O(A_0,E_0)$, that
 $\Delta {\hat O}$ be small compared to typical classical values of the
function $O$ and that the usual correspondence between commutators and 
Poisson brackets holds for expectation values in quasi-classical
states.

We end this section with a few technical remarks. If for every $L$,
${\hat O}_L$ is a bounded self adjoint operator on ${\cal H}_L$ then
using 
Lemma 1, section 4.4 of \cite{al}, it can be verified that ${\hat O}$
is an essentially self adjoint operator on the dense domain of finite
linear combinations of spin networks in 
$L^2(\overline{{\cal A}/{\cal G}}, d\mu_H)$.

However, typically, the operators ${\hat O}_L$ of interest are
(unbounded)
densely defined 
symmetric operators on ${\cal D}_L$. Then it is straightforward
to see that ${\hat O}$ is a densely defined symmetric operator on the 
dense domain of finite linear combinations of spin network states in 
$L^2(\overline{{\cal A}/{\cal G}}, d\mu_H)$.

\section*{3. `Magnetic flux' operators.}

Holonomies serve as natural candidates for the classical functions `$O$' 
of section 2, in as much as the connection degrees of freedom are concerned.
>From the considerations of section 2, we restrict attention to holonomies 
along loops which lie on the lattice associated with a quasiclassical state.
The classical metric being approximated endows every such loop with a size.
Clearly, it does not make sense to require that holonomies along Planck size
loops display classical behaviour; it is only for loops of size much larger 
than the Planck scale, that we expect classical behaviour. Hence, we may 
further restrict our attention to holonomies along such ``large''
loops.

A different set of operators than the large loop holonomies is suggested 
by the lattice structure. These operators are the magnetic flux operators of 
lattice gauge theory which measure the non-abelian magnetic flux through the
plaquettes of the lattice. They are defined in the natural way via holonomies
along the plaquettes \cite{creutz}. 

Since a single plaquette is typically of Planck size, we shall refer to the 
magnetic flux through a plaquette as the `microscopic' magnetic flux.
Clearly, the microscopic magnetic flux is not of direct relevance to the 
classical limit. It is only `macroscopic' operators associate with 
`macroscopic' length scales (i.e. length scales far above the Planck scale)
that are relevant to the classical limit. The utility of the microscopic 
magnetic flux  (or equivalently, the holonomy along a `microscopic' loop) 
is that it serves as the lattice approximant to the curvature of the 
connection - the curvature is approximated on the lattice by the 
flux through a plaquette divided by the plaquette area. Many physically
interesting functions can be constructed from the curvature (for e.g.
$D({\vec N})=\int_{\Sigma}  N^a {\tilde E}^a_iF_{ab}^i$, 
where $N^a$ is a vector field
and $F_{ab}^i$ the curvature of the connection) and thus, admit lattice 
approximants built out of  microscopic fluxes.

It turns out, as we show in section 3.1, that because of the differences
in their algebraic properties, it is simpler to use the flux operators rather
than the holonomies along macroscopic loops, to analyse the classical limit.
Moreover, as discussed in section 3.2, the consideration of flux-based 
macroscopic operators suggests a general strategy to build states in which 
these operators have low relative fluctuations. For these reasons we shift 
focus from the holonomies of macroscopic loops to flux based macroscopic 
operators in our explicit constructions of section 4. As we shall see in 
section 5, the strategy discussed in section 3.2 is not entirely 
successful; nevertheless this strategy springs from an interesting idea 
and, among other things, this work is devoted to examining it in detail.

\section*{ 3.1. Algebraic properties of holonomies vs fluxes.}
The holonomies and fluxes have very different algebraic properties.
Fluxes are associated with 2d surfaces and and are {\em additive}.
The flux through the union, $S$,  of disjoint surfaces $S_I, \;I=1..M$
is the sum of the 
fluxes through each of the surfaces,
\begin{equation}
\int_{S}F_{ab}^i = \sum_{I=1}^M\int_{S_I}F_{ab}^i .
\end{equation}
Here $F_{ab}^i$ is the curvature of the connection pulled back to the
relevant 2 surface and $i$ is some fixed internal $SU(2)$ direction.
Equivalently, defining the flux $\Phi^i(S) = \int_{S}F_{ab}^i$,
\begin{equation}
\Phi^i(S)=\sum_{I=1}^M\Phi^i(S_I)
\label{eq:algflux}
\end{equation}
In contrast holonomies are associated with 1d loops and are 
{\em multiplicative}. Thus if 
$\gamma := \gamma_1 \circ \gamma_2....\circ\gamma_N$ is the loop
composed of the loops $\gamma_I, I=1..N$, 
\begin{equation}
H_{\gamma}(A) = \prod_{I=1}^NH_{\gamma_I}(A)
\label{eq:alghol}
\end{equation}
where the product signifies group multiplication.

Thus, (\ref{eq:algflux}) determines the flux through large surfaces in terms of
small surfaces which combine to form the large surfaces and 
(\ref{eq:alghol}) determines the holonomy of a composite 
loop in terms of the holonomies of the loops which compose it.

By definition, (\ref{eq:algflux}) also holds for the quantum flux operators
and hence for their expectation values. Thus, the expectation values of the 
fluxes through small surfaces determine the expectation value of
the flux through the large surface via the quantum version of 
(\ref{eq:algflux}). This simplifies the construction of quasiclassical 
states since it suffices to restrict attention to a smaller ``basis'' set of
surfaces from which all surfaces of interest can be composed. Similar
considerations hold for gauge invariant flux based macroscopic operators.

In contrast, although (\ref{eq:alghol}) also holds for the holonomy operators,
it does not necessarily hold for their expectation values due to 
quantum fluctuations. In fact, as we show below, if (\ref{eq:alghol})
is {\em imposed} as a relation between expectation values in a quantum state,
that state cannot be quasiclassical. This complicates the construction of 
quasiclassical states; since we cannot restrict attention to a smaller 
``basis'' set of loops, the holonomies have to be approximated all at once.
 It is in this sense that it is easier to use fluxes than holonomies.

We now prove our claim regarding the holonomy expectation values.

On $L^2({\overline{\cal A}}, d\mu_H)$ 
define the bounded self adjoint operators
\begin{equation}
{\hat x}^0_{\alpha}:= {\hat T}^0_{\alpha}, \;\;\;\;\;\;\;\;\;\;
{\hat x}^j_{\alpha}:= {i\over 2} Tr ({\hat H}_{\alpha}\sigma^j),
\end{equation}
where $\sigma^j$ are the $2\times 2$ Pauli matrices. Since 
$H_{\alpha}(A) \in SU(2)$,
\begin{equation}
\sum_{\mu=0}^3 ({\hat x}^{\mu})^2 = det {\hat H}_{\alpha}=1.
\label{eq:det}
\end{equation}
For any state in $L^2({\overline{\cal A}}, d\mu_H)$,
\begin{equation}
det <{\hat H}_{\alpha}> =\sum_{\mu=0}^3 <{\hat x}^{\mu}>^2.
\label{eq:detmean}
\end{equation}
>From (\ref{eq:det})
\begin{equation}
det <{\hat H}_{\alpha}> = 1- \sum_{\mu=0}^3 (\Delta {\hat
x}^\mu_{\alpha})^2.
\label{eq:detdelta}
\end{equation}
>From (\ref{eq:detmean}) and (\ref{eq:detdelta}),
\begin{equation}
0 \leq det <{\hat H}_{\alpha}> \leq 1- (\Delta {\hat T}^0_{\alpha})^2
\label{eq:dett0}
\end{equation}
Let $\gamma_I, I=1..N$ be a set of $N$ loops such that their
composition is 
the loop $\gamma$. 
Thus, $\gamma := \gamma_1 \circ \gamma_2....\circ\gamma_N$.

Let $(A_{0a}^i, {\tilde E}^a_{0i})$
be the classical data to be approximated. 
Let $\epsilon >0 $ be a physically reasonable 
lower bound on the attainable uncertainty in the measurement of 
the ${\hat T}^0_{\gamma_I}, \; I=1..N$. Thus,
\begin{equation}
\Delta {\hat T}^0_{\gamma_I} \geq \epsilon >0,
\label{eq:deltat0}
\end{equation}
Since 
${ H}_{\gamma}(A_0) = \prod_{I=1}^N { H}_{\gamma_I}(A_0)$ we impose
that 
%BELOW I CHANGED THE ORIGINAL ``='' TO SIGNS OF APPROX
\begin{equation}
<{\hat H}_{\gamma}> \approx \prod_{I=1}^N <{\hat H}_{\gamma_I}>.
\label{eq:ruff0}
\end{equation}
\begin{equation}
\Rightarrow det <{\hat H}_{\gamma}> \approx 
\prod_{I=1}^N det <{\hat H}_{\gamma_I}>.
\label{eq:ruff}
\end{equation}
Since $L^2(\overline{{\cal A}/{\cal G}}, d\mu_H) \subset
L^2(\overline{{\cal A}}, d\mu_H)$, we can use 
(\ref{eq:dett0}) to get 
\begin{equation}
 det <{\hat H}_{\gamma}> \leq 
\prod_{I=1}^N 
1- (\Delta {\hat T}^0_{\gamma_I})^2 .
\label{eq:ruff7}
\end{equation}
{}From (\ref{eq:detmean}), (\ref{eq:deltat0}) and (\ref{eq:ruff7})
\begin{equation}
|<{\hat T}^0_{\gamma}>|^2 < det <{\hat H}_{\gamma}> < (1-\epsilon)^N. 
\label{eq:t0=0}
\end{equation}
Since $\epsilon$ is independent of $N$, 
clearly, for sufficiently large $N$, the above equation implies that 
$|<{\hat T}^0_{\gamma}>| << 1$. 
For generic $A_{0a}^i$ there is no reason for 
the classical variable $T^0_{\gamma}(A_0)$ to be small. 
So if we assume that the classical connection of interest is such that
\begin{equation}
T^0_{\gamma}(A_0) \sim O(1),
\label{eq:assume1}
\end{equation}
then (i) is clearly violated for the loop $\gamma$ 
because $|<{\hat T}^0_{\gamma}>-T^0_{\gamma}(A_0)|$ is {\em not} much
less
than unity.
%Hence, the candidate state is not quasi-classical.

For loops of macroscopic size, we obtain   
rough estimates for $N$ and $\epsilon$ as follows.
Quantum gravitational fluctuations are not expected to be significant
well above the Planck scale. So for the purposes of the 
gravitational interaction alone,  energy scales of up to a 
few hundred $Gev$ (or equivalently length scales larger than 
 $10^{-16} cm$) can safely be considered as `classical'. A macroscopic
size
surface
of the order of $100 m^2$ contains 
the loops $\gamma_I, I= 1..N$, where each $\gamma_I$
encloses a `classical' size area of the order of $10^{-32}cm^2$. 
Thus $N$ is of the order of $10^{38}$.
Even if $\epsilon$ is chosen as small as $10^{-34}$, we obtain
\begin{equation}
|<{\hat T}^0_{\gamma}>| \leq (1- 10^{-34})^{10^{38}}
 \; \sim \; e^{-10^4} \sim 0,
\label{eq:estimate}
\end{equation}
which clearly violates (i) for classical 
connections which satisfy (\ref{eq:assume1})!

One way of arriving at a physically motivated choice for $\epsilon$ 
is as follows. In addition to the loops $\gamma_I$, consider a set of
surfaces 
$S_J, \;J=1..N$, each of classical size $10^{-32}cm^2$ such that each 
$\gamma_I$ transversely intersects $S_I$ exactly once. Then, choosing
the orientation of $S_I$ to be in the direction of $\gamma_I$, 
and denoting the area of the surface $S_I$ by $A_{S_I}$,
we have
\begin{equation}
\{T^0_{\gamma_I}, A_{S_I} \}
=-\iota G_0
{i\over 2} Tr (H_{\gamma_I}\sigma^i) n_i,
\end{equation}
where the right hand side is evaluated at the point of intersection
between
the loop $\gamma_I$ and the surface $S_I$.
$n_i$ is defined as follows. Let $E^a_i :={{\tilde E}^a_i\over
\sqrt{q}}$
where $q$ is the determinant of the metric 
constructed from the triad. 
Let 
 $n_a$ be the
unit normal to the surface $S_I$ defined by  this metric. Then 
$n_i:=n_a E^a_i$. Thus $n_i n^i =1$ and we expect  that for a large 
class of connections (with less than Planck scale curvature
and which also satisfy 
(\ref{eq:assume1})) and  triads, it should be true that 
\begin{equation}
\{T^0_{\gamma_I}, A_{S_I} \}= -\iota G_0
{i\over 2} Tr (H_{\gamma_I}\sigma^i) n_i  \sim \iota G_0 O(1).
\label{eq:poisson}
\end{equation}
Note that if the above equation holds, then $Tr (H_{\gamma_I}\sigma^i)$
is of order unity. This implies that the curvature of the connection, 
$F_{ab}^i$, in physically reasonable coordinates is of the order of 
$10^{32}cm^{-2}$ which is still, for purposes of quantum gravity,
classical.

For quasi-classical states we expect that the Poisson bracket to quantum
commutator correspondence  holds in the sense of expectation value so
that
\begin{equation}
i\hbar 
\{T^0_{\gamma_I}, A_{S_I} \}\sim <[{\hat T}^0_{\gamma_I}, {\hat
A}_{S_I}]>.
\label{eq:pbtocom}
\end{equation}
Combining (\ref{eq:poisson}) with (\ref{eq:pbtocom}) with the
uncertainty 
principle for $\Delta {\hat T}^0_{\gamma_I}$, $\Delta {\hat A}_{S_I}$
we get
\begin{equation}
\Delta {\hat A}_{S_I}\Delta {\hat T}^0_{\gamma_I} \sim \iota l_{0P}^2 
\label{eq:estimateuncert}
\end{equation}
Let us  assume, to be conservative,
 a huge uncertainty in the measurement of area
\footnote{Note that our estimates are in the context of a thought
experiment
in which the only quantum effects are from the gravitational 
interaction. In practice, it would of course be almost impossible to
directly make the appropriate measurements, due, in part, to the 
quantum nature of any interaction used in the measuring process.} 
equal to  $10^{-32}cm^2$
and set $\iota l_{0P}^2$ to be of the order of  the Planck area 
(the latter is consistent
with the black hole entropy calculations of \cite{entropy}).
Then from (\ref{eq:estimateuncert})
$\epsilon = {10^{-66}\over 10^{-32}}= 10^{-34}$.

Finally, we note that 
(\ref{eq:ruff0}) mirrors the relations (\ref{eq:alghol}) 
between classical holonomies.
Since classical holonomies are not gauge invariant objects, it is
necessary to
extend our arguments to the gauge invariant context of traces of
holonomies. We do this in appendix A1 by using Giles' (re)construction 
\cite{giles} of 
holonomies from their traces.

This completes our discussion as to why it is technically simpler to use 
fluxes as opposed to holonomies.

\section*{3.2.  A general strategy for low fluctuations based on flux 
operators}
In this section we describe a general strategy to obtain low relative 
fluctuations  of flux based `macroscopic' operators. This strategy is 
patterned on the mechanism for low relative 
fluctuations in statistical mechanics.
In the statistical mechanics description of thermodynamic systems, there 
are `$N$' weakly correlated degrees of freedom, $N$ being very large.
Mean values of macroscopic quantities typically go as $N$ times some 
microscopic quantity whereas the relative fluctuations about the mean go 
as ${1\over \sqrt{N}}$. It is the poor correlation between the 
degrees of freedom that is responsible for such low relative fluctuations.

How can we use this mechanism for low relative fluctuations in the 
context of our proposal? Recall from section 2 that the lattices of physical
interest associated with quasiclassical states have links which are of the 
order of the Planck length. A `macroscopic' lattice operator, ${\hat O}_L$,
associated with a classical function $O$ is typically the sum over 
`$N$'microscopic operators ${\hat O}_{I_L}$. The index ${I_L}$ 
typically ranges over all the plaquettes/cells in a macroscopic volume.
Since the cells are of Planck size, $N$ is very large.
This raises the possibility of constructing states with
${1\over {\sqrt{N}}}$ relative fluctuations in the measurement of
${\hat O}$. We indicate how this 
could happen below and show that it is possible to construct such 
 states in the next section.
 
>From (\ref{eq:pl}) and  (\ref{eq:defo}) it is easy to see that 
\begin{equation}
{\hat O}^2 = \sum_L {\hat P}_L {\hat O}_L {\hat P}_L{\hat O}_L {\hat
P}_L . 
\end{equation}
%Define 
%\begin{equation}
%{\hat O^{\prime}}^2 := 
%\sum_L {\hat P}_L {\hat O}_L^2 {\hat P}_L . 
%\end{equation}
It can be checked that 
\begin{equation}
<\sum_L {\hat P}_L {\hat O}_L^2 {\hat P}_L > \geq <{\hat O}^2>.
\end{equation}
\begin{equation}
\Rightarrow 
(\Delta^{\prime} {\hat O})^2 := 
<\sum_L {\hat P}_L {\hat O}_L^2 {\hat P}_L>- <O>^2
\geq (\Delta {\hat O})^2 .
\label{eq:deltaprime}
\end{equation}
It can be verified that $\Delta^{\prime} {\hat O}$ evaluated in
the quasi-classical state based on the lattice $L_0$ is given by 
\begin{equation}
\Delta^{\prime} {\hat O} = \Delta {\hat O}_{L_0}.
\end{equation}
But 
\begin{equation}
{\hat O}_{L_0}= \sum_{I_{L_0}=1}^{N}{\hat O}_{I_{L_0}}.
\label{eq:sumol}
\end{equation}
Then, if the ${\hat O}_{I_{L_0}}$ are sufficiently uncorrelated in the 
state, we have for $I_{L_0}\neq J_{L_0}$ that
\begin{equation}
<{\hat O}_{I_{L_0}}{\hat O}_{J_{L_0}}>\approx 
<{\hat O}_{I_{L_0}}><{\hat O}_{J_{L_0}}>.
\end{equation}
Then (\ref{eq:sumol}) implies that 
\begin{equation}
(\Delta {\hat O}_{L_0})^2 \approx \sum_{I_{L_0}=1}^{N}
(\Delta {\hat O}_{I_{L_0}})^2.
\end{equation}
Typically, we expect
$<{\hat O}_{I_{L_0}}>$ and $\Delta {\hat O}_{I_{L_0}}$ to be 
of order 1 times
some microscopic (in general, dimension-full) constant and 
$<{\hat O}>= \sum_{I_{L_0}=1}^{N}<{\hat O}_{I_{L_0}}>$ to be 
 of order $N$ times the same constant
\footnote{Unfortunately, as we  shall see in section 5, 
our strategy is not entirely 
successful because this expectation is not quite true for the operators 
and the states that we 
examine in section 4.}. Then we get 
\begin{equation}
{\Delta {\hat O}\over <{\hat O}>} 
\leq {\Delta^{\prime} {\hat O}\over <{\hat O}>}
={\Delta {\hat O}_{L_0}\over <{\hat O}_{L_0}>} \approx {1\over
\sqrt{N}}.
\end{equation}

In section 4 we shall examine  some classical functions and their flux-based 
lattice approximants, and apply the strategy of this section to construct 
states with low relative fluctuations of the corresponding operators.

\section*{4 .Kinematical $2+1$ gravity}\label{2+1}
In subsections 4.1-4.3, we explore our ideas in the context of 2
spatial 
dimensions. In 4.1 we define some macroscopic functions and construct
their 
quantum analogs in accordance with (\ref{eq:defo}). In 4.2 we construct
candidate quasi-classical states. In 4.3 we show that the relative 
fluctuations of the macroscopic operators defined in 4.1 can go as 
${1\over \sqrt{N}}$ in accordance with the ideas of section 3.

Unfortunately, for the reason mentioned in footnote 8, it is difficult to keep 
the scale of the  fluctuations 
%(as opposed to the relativefluctuations)
of these operators smaller than the typical scale of the corresponding 
classical quantities. Hence, not all of our ideas are successfully 
implemented. A discussion presenting ways in which 
our states may be modified, or our strategy refined is also contained in 
section 5.

In subsection 4.4 we indicate how to generalize our constructions to 
three spatial
dimensions. We note that the same difficulties with the scale of the 
fluctuations arise there, too, and hence our construction of 
quasi-classical states is not yet satisfactory. 

\subsection*{4.1 The macroscopic observables}\label{obs}

 In two spatial dimensions the phase space variables are 
a densitized triad and a $SU(2)$ connection 
$({\tilde E}^a_i, A_a^i)$ \cite{thiemann2+1} where $i$ is an $SU(2)$
Lie algebra index and $a$ is the spatial index. The metric is
constructed 
from ${\tilde E}^a_i$ through ${\tilde E}^a_i{\tilde E}^{bi} = q
q^{ab}$.
In two dimensions the spatial geometry is determined if the lengths of
all curves
in the 2-manifold are specified. Moreover, for non-degenerate ${\tilde
E}^a_i$
(i.e. for ${\tilde E}^a_i$ which define non-degenerate 2 metrics), the 
information in the curvature $F_{ab}^i$ of the connection is coded in
the 
local expressions ${\tilde E}^a_iF_{ab}^i$ and 
$\epsilon^{ij}_k{\tilde E}^a_i{\tilde E}^b_jF_{ab}^k$. 
Hence the classical functions of interest are the length of an
arbitrary 
curve `$c$', $l(c)$, the `vector constraint',
 $D({\vec N})=\int  N^a {\tilde E}^a_iF_{ab}^i$,
and the `scalar constraint', 
 $S(N)=\int \NN \epsilon^{ij}_k{\tilde E}^a_i{\tilde E}^b_jF_{ab}^k$
where 
$N^a$ is an arbitrary vector field and $\NN$ is a density -1 scalar.

The corresponding operators are constructed as follows.
The length operator can be constructed independent of the strategy of 
section 3, in the same fashion as the area operator in 3d.
%In  two dimensions the spatial 
%geometry is completely determined 
%by the length operators. We will assume that 
The eigenstates of the length operator ${\hat l}(c)$ 
are the spin network states
and their eigen values have a contribution of 
$\lambda_j = 2 l_P \sqrt{j(j+1)}$ for every intersection 
of the curve $c$ and a link of the spin network colored by $j$.
Here $l_P := \iota l_{0P}$.
Note that in the language of section 3,  
this operator induces length operators $\hat{l}_L(c)$ in any lattice
$L$. 

The two sets of connection dependent operators can be defined first on 
a lattice $L$ and then promoted to genuine operators on 
$L^2({\overline{\cal A/G}}, d\mu_H)$ through (\ref{eq:defo}).

\begin{equation}
\hat{D}_L(\vec{N}) = \frac{1}{4} 
\sum_{v, p=l_{p1} \wedge l_{p2}\to v} 
\hat{F}(p) \cdot (\hat{E}(v,l_{p1}) N(v,l_{p2}) - 
\hat{E}(v,l_{p2}) N(v,l_{p1}) )+\rm{H.T.}  
\label{D}
\end{equation}
where the sum runs over all vertices and all plaquettes that
contain 
each given vertex (at vertex $v$ the orientation of plaquette $p$ is 
given by an ordered pair of links $l_{p1} \wedge l_{p2}$), 
$\hat{F}^i(p) = {-i\over 2}{\rm Tr}(H(p) \sigma_i)$ 
($H(p)$ is the holonomy around 
plaquette $p$) and $\hat{E}(v,l)$ 
acts as a left invariant vector field (multiplied by a factor of $l_P$)
 on functions depending on the
holonomy
along the link $l$ oriented away from vertex $v$ . `H.T.' refers to
Hermitian transpose.

%It is important to clarify the meaning of our notation: 
$\hat{E}(v,l)$ can be interpreted
 as the triad operator smeared over a line 
transverse 
to the link $l$, but not crossing $l$ in the center but at $v$. 
$\hat{F}(p)$ contains the information of the curvature smeared in the 
plaquette (plus higher order terms in the curvature that are not small 
in general). 
%Because of this interpretation, in certain regimes of the 
Thus, $\hat{E}$ and $\hat{F}$ are related to the triad and the 
curvature times factors of the lattice spacing 
$a_g$, measured by the macroscopic metric induced by the length
operator 
in our state. 
%Thus, the expectation values of this family of microscopic operators
%could 
%approximate the 
 Equation (\ref{D}) provides a discretization of the classical
%/macroscopic functions known as 
 vector 
constraints if the vector
field $N^a$ and the collection of weights
assigned to 
the lattice links are related by 
$N(v,l_{p1})\hat{l_{p1}} + N(v,l_{p2})\hat{l_{p2}}= 
\frac{1}{a_g} \vec{N}(v)$, with 
$\hat{l_{p1}}$, $\hat{l_{p2}}$ being unit vectors in the
direction of two of the links starting at $v$ and forming a
right-handed
basis.% 
\footnote{We assume that the vertex is four valent and 
formed by the intersection of two smooth
curves; in this way, the definition does not depend on the choice of
links to
form the basis. }
A definition of $\hat{D}_L(\vec{N})$ which corresponds to 
the classical function $D(\vec{N})$ for states with 
arbitrary valence would be more cumbersome to write. Since most of the 
vertices in the states that we will construct are four valent the 
expression (\ref{D}) for $\hat{D}_L(\vec{N})$ is good enough for our
purposes.

The other family of operators is defined by 

\begin{equation}
\hat{S}_L(\NN) = \frac{1}{4} 
\sum_{v, p=l_{p1} \wedge l_{p2}\to v} 
{\NN}_L(v) \hat{F}^i(p) \hat{E}^j(v,l_{p1})  \hat{E}^k(v,l_{p2})
\varepsilon_{ijk} +\rm{H.T.} 
\label{}
\end{equation}
For this family of operators, the expectation values 
(on states with mostly four valent vertices) will approximate 
the classical functions known as the scalar constraints if the scalar
of 
density weight $-1$ labeling the functions is related to the collection
of weights assigned to the vertices 
by the relation ${\NN}_L (v)= \frac{1}{a^2_g} \NN (v)$. 

\subsection*{4.2 States with  ${1\over \sqrt{N}}$ 
relative fluctuations }\label{qstates}

In this section we display candidate quasi-classical states which 
provide a realization of our idea of ${1\over \sqrt{N}}$ relative
fluctuations.
% as mentioned earlier, and discussed in section 5, our 
%constructions  are still unsatisfactory. Nevertheless, we display them 
%because their failure is due to a robust obstruction, not to a slight 
%error in the design. As we discuss in section 5, the obstruction is 
%due to an uncertainty relation, and 
%the choice of small spins in the construction of the 
%states (which is a general preconception in the community; see for 
%example \cite{weave}). 
As mentioned earlier and discussed in section 5, the states which we 
construct are not completely satisfactory quasiclassical states.
Neverthless, we present the construction of the candidate
quasiclassical states in detail in the 
 hope that this may
fuel future efforts towards modifying 
our present strategy appropriately.

We shall display candidate 
quasi-classical states approximating homogeneous geometries and
connections. 
This family of states includes, for example, states that generate 
expectation values approximating 
Euclidean metrics and 
 flat connections on a torus, as well as states which  
approximate round spheres
with constant curvature $SU(2)$ connections on them%
\footnote{
We remind our reader that the macroscopic observables that we are
studying 
now are of local character and therefore two gauge inequivalent classical flat 
connections would appear indistinguishable to our ``magnetic flux
type'' 
observables. 
}. 

To make the macroscopic geometry 
(locally) isotropic, the physical lattice
prescribed 
by the state will cover space with domains with the connectivity of a
regular
square lattice; these domains will be 
separated by narrow bands. We demand the distribution of
orientations of the regular domains to be isotropic. 
The dominant contributions to any macroscopic
observable will be those coming from the interior of the regular
domains, 
and many domains will be involved in any macroscopic measurement. 
Thus, macroscopic
observables will lose track of the connectivity of the lattice which will
only be
obvious at the micro-scale.

Our lattice should be composed by regular domains of typical size 
$D > > > l_P$  and have a linear density of links $\rho_l =
\frac{k}{l_P}$. 
This is the density of intersections of the lattice links with any
curve which
wiggles only at the macroscopic scale (technically, its radius of
curvature
should be macroscopic). With this linear density of links the density
of
plaquettes is $\rho_p = (\frac{k \pi}{4 l_P})^2$. We will later show
that 
$k=\frac{2}{\sqrt{3}}$ is the correct value of this parameter, given
the
form of the states described below.

The states will be constructed as products; to each regular domain we 
will assign a factor and a separate factor will be assigned to the 
region between the regular domains. The factors assigned to regular 
domains will also be constructed as products. 
Taking advantage of the regularity, the interior of 
the domains are divided into 
black and white plaquettes in an alternate fashion. 
In the chess-board-like geometry of the interior
of the domains we will assign factors of the wave function only to the black
plaquettes 
asking that the color $n=2j=0$ does not appear in the spin
network 
decomposition of any of the factors. 
Due to the alternate plaquette geometry,
the color
assigned to the links in a spin network decomposition of the state 
would be exactly the one 
coming from its only black plaquette neighbor. 
In this way our quasi-classical state will
provide a physical lattice. 
It will be important that the spin network
decomposition of the state does not acquire any zero color in 
the region between the regular domains 
to make sure that the state does prescribe a physical lattice and
not a
collection of separate domains. A technique to fit the domains together 
will be described after the contributions  from the interior of the
domains
are explained.

As we mentioned earlier, the factor of the wave function assigned to 
a domain is a product of factors associated to plaquettes 

\begin{equation}
\psi_D = \prod_{p \in Bl} \psi_p
\label{34}
\end{equation}
\noindent 
where $Bl$ contains alternate plaquettes.
Since
there are many more plaquettes in the interior of the domains than in
the 
region between domains, to approximate any macroscopic observable we
need to 
adjust only the factors associated to interior plaquettes. Furthermore,
since
we will illustrate our construction with a state approximating a 
homogeneous
geometry and a homogeneous connection, all the factors $\psi_p$ from
the interior
plaquettes can be taken equal. We choose 

\begin{equation}
\psi_p = \cos{\theta} \phi_{n=1} + \sin{\theta} \phi_{n=2}  \quad .
\label{}
\end{equation}
where $\phi_i(A)$ is the trace of the holonomy around plaquette $p$ in 
the spin $\frac{i}{2}$ representation. 
Other choices of $\psi_p$ are possible;
%(we could use, for example, 
%gaussian 
%factors as suggested by Corichi and Reyes \cite{Corichi-Reyes} 
%in a different context); 
we chose the simplest states that defined a physical lattice and had 
small spins.

Let us now describe the assignment of factors of the wave function to
the regions of the lattice 
that do not belong to the domains described earlier. 

To simplify our work we will restrict the geometry of the lattices that 
we consider. First, we concentrate on the boundary of the regular domains.
The boundary of the regular domains will be composed only of black 
plaquettes (one may construct these kind of geometries by erasing the 
boundary links of the white plaquettes in the boundary). 
In addition, we 
will only consider geometries where the black plaquettes 
in the boundary of the regular domains share at least two vertices with 
the black plaquettes in the domain, and if one 
of these black plaquettes shares only two vertices with the interior 
plaquettes this plaquette must be triangular (the plaquettes in the 
interior of the domains are all square plaquettes, but in the boundary 
we allow also triangular plaquettes). 
In these geometries it follows that all the 
plaquettes having a link in the boundary 
are black and that these boundary plaquettes have at most 
one vertex that is not 
shared by any plaquette in the regular domain. We will call these
vertices 
black vertices. Apart from these vertices, in 
the boundary of the regular domains, there are vertices that are shared 
by interior plaquettes. We will call these vertices, white vertices. 

We will take these boundary vertices as data and construct the rest of
the 
lattice by filling the gaps in between the domains in a way that lets us 
assign a simple factor of the wave function to this ``in between'' region of 
the lattice. 

At a bigger scale 
we can use the regular domains as cells of a 
latticization
\footnote{
The 1-skeleton of cellular complex with the topology of $\Sigma$.
} 
of the surface $\Sigma$.
Neighboring domains are separated by bands 
(analog of links) and these bands meet in rotaries (analog of
vertices). 
For convenience, in the 
lattices that we will consider the bands and rotaries will have 
no internal vertices, and the rotaries will have no internal links. In
other 
words, the rotaries are simply cells whose links are 
boundary links of the bands or boundary links of the regular 
domains. 
On the other hand, the bands have interior links, but the 
interior links of each band are restricted 
to form a closed curve $\gamma_B$ 
joining black vertices (either joining black vertices from the same 
domain or joining black vertices of neighboring domains). 
See the figure.

\vskip .9cm
\centerline{\epsfxsize=6.5cm
\epsfysize=3.5cm
\epsffile{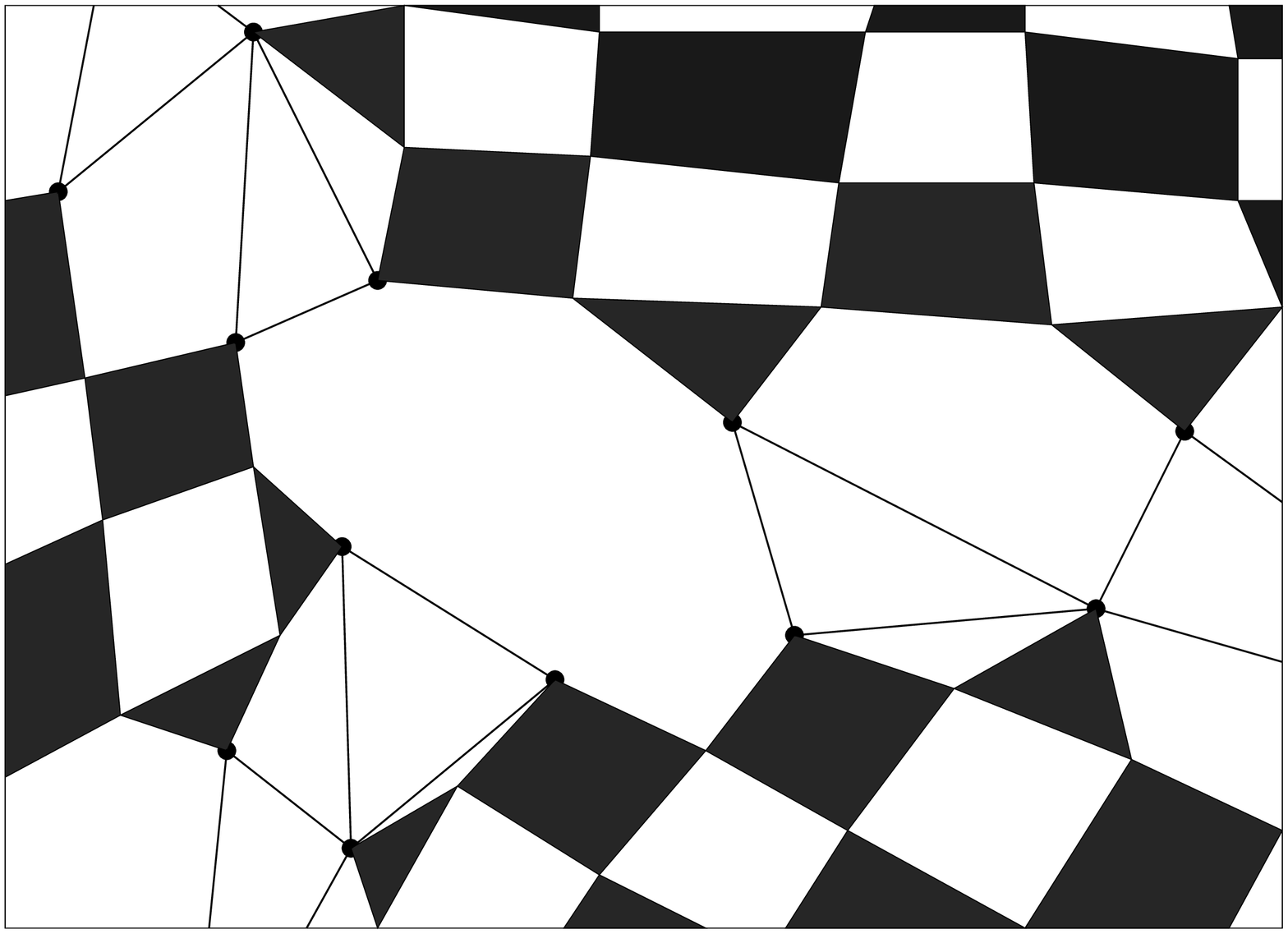}}

%\caption{ 
{\small {\bf Fig. 1}
We show a region of the lattice that is in the boundary between three 
regular domains. The domains have square lattice connectivity in 
their interiors and we assign factors of the 
wave function to the black plaquettes in the chess-board geometry of the 
interiors. Separate factors are assigned to each of the 
bands that serves as boundary between two regular domains; these factors 
are spin networks of color one whose graph $\gamma_B$ is 
drawn joining the black vertices of the figure without retracing any line. 
}
\vskip .7cm

Due to the connectivity of its interior links, we can assign to a 
band a factor of the wave function which is simply the spin network 
determined by its graph and the color $n=1$, 
$\psi_B = \psi_{\gamma_B, n=1}$. 
Since the links of a band 
join only black vertices, when we multiply the band factors with the 
domain factors, the spin network decomposition of the state will not 
have color zero in any link. That is, our proposed wave function 
\begin{equation}
\psi = [\prod_D \psi_D] [\prod_B \psi_B]
\label{}
\end{equation}
\noindent 
defines a physical lattice which encodes the topology of $\Sigma$.

It is clear that a space manifold with arbitrary 
topology can be covered by a lattice composed by disconnected domains 
with trivial topology (whose interiors have the connectivity of a
regular 
square lattice and required link density) and 
joined by narrow band regions where the lattice does not 
need to posses any regularity. Thus, from the classical data of a
Euclidean 
torus with a flat $SU(2)$ connection we can construct our 
candidate quasi-classical 
states based on the required lattice, and 
analogously from the classical data of a 
round sphere with a constant curvature connection we can construct 
candidate quasi-classical states.

\subsection*{4.3 Expectation values, 
fluctuations and correspondence}\label{conditions}

Given a macroscopic curve we want to calculate the expectation value of
its
length $< \hat{l}_L(c) >$. The calculation is easy. In two
dimensions the eigenstates of the length operator are spin network
states 
and their eigen values have a contribution of 
$\lambda_n = \frac{1}{2} l_P \sqrt{n(n+2)}$ 
for every intersection with the curve.
According to our conventions, the total number of intersections is 
$\frac{k}{l_P} l_g(c)$, 
where $l_g(c)$ is the length of the curve. 
Now we will make two approximations; we will assume that
every
plaquette which intersects $c$ intersects it twice and that the
parameter 
$\theta$ is small (because, as we will see, it is linked to the
contribution of one plaquette to the curvature of the connection).  
In this way we get  
\begin{eqnarray}
< \hat{l}_L(c) > &=& 2 \sum 
\lambda_{n=1} \cos^2{\theta} + \lambda_{n=2} \sin^2{\theta} \nonumber
\\
&\approx& \lambda_{n=1} \frac{k}{l_P} l_g(c)
\label{}
\end{eqnarray}
where the sum runs over all plaquettes that intersect the curve $c$. 
>From this formula we deduce $k =
\frac{l_P}{\lambda_{n=1}}=\frac{2}{\sqrt{3}}$.

Now we calculate the expectation values of the connection related
observables. 

The 
connection measuring operators are a sum of terms. 
%they exhibit some degree of locally. 
%In particular we have that  
The dominant contributions to the expectation values 
come from the interior of the regular domains. Since
%can be calculated doing only a few integrals and 
each term in the sum only affects a few factors in the wave
function, the contributions can be easily calculated.
%In addition, to make things even simpler, 
Since we 
deal only with homogeneous wave functions, the calculations simplify even more.
 If we had allowed 
the parameter $\theta$ to be a function of the plaquette the basic 
mechanism of our proposal would still work; we would be able to 
approximate many more classical configurations, but the calculations 
would not be as simple.

It turns out that the alternate plaquette nature of the 
wave function makes \\
\noindent $<\hat{D}(\vec{N})> = 0$ for all 
constant shifts. This would be true even if the parameter $\theta$ 
were a function of the plaquette. To see this, it is convenient 
to rewrite the action of the vector fields ${\hat E}^i(v,l_x)$ 
in a way that is tailored to act on wave functions that are 
products of plaquette factors. 
When acting on fucntions of type (\ref{34}), 
$\hat{E}^i(v,l_x) = 
\frac{i l_P}{2} [ \hat{L}^i(p) - \hat{L}^i(p,-\hat{y})]$, 
where $\hat{L}^i(p), \hat{L}^i(p,-\hat{y})$ are the left
invariant 
vector fields acting on functions of the group assigned to the 
plaquette $(p)$ and the neighbor of $(p)$ in the $(-\hat{y})$
direction respectively. 
Due to the alternate plaquette geometry, the only terms that do not
vanish 
in the expectation value are the ones containing the factor 
$\hat{F}(p) \cdot \hat{L}(p)$. The result is 

\begin{equation}
< \hat{D}(\vec{N}) > 
= \sum_{p \in Bl}
C_0 {\rm Div}N(p) 
\label{}
\end{equation}
where, in the plaquette $(p)$ defined by the vertices 
$(v, v + l_{p1}, v + l_{p1} + l_{p2}, v + l_{p2})$, 
\begin{eqnarray}
{\rm Div}N(p) = 
N(v,l_{p1}) + N(v,l_{p2}) - 
N(v+l_{p_1},l_{p1}) + N(v+l_{p_1},l_{p2}) - \nonumber\\
N(v+l_{p_1}+l_{p2},l_{p1}) - N(v+l_{p_1}+l_{p2},l_{p2}) + 
N(v+l_{p2},l_{p1}) - N(v+l_{p2},l_{p2}) \nonumber
\end{eqnarray}
and $C_0$ is calculated for any interior plaquette $(p)$ as
\begin{equation}
C_0 = \frac{i l_P}{2} 
< \hat{F}(p) \cdot \hat{L}(p) - H.T. > 
\end{equation}
%where the calculation of the expectation value involves only a few 
%integrals easily done using 
%$\hat{F}(p)\cdot \hat{L}(p) \psi_{p,n} 
%= c_{n-} \psi_{p,n-1} + c_{n+} \psi_{p,n+1}$.
%In our case $c_{1+}= -\frac{1}{2}$ and $c_{2-}= -\frac{3}{4}$ and, 
For small $\theta$, 
$C_0\approx - \frac{5}{4}\theta l_P$.

For similar reasons, the expectation value of $\hat{S}(\NN)$ 
simplifies greatly  and we get
\begin{equation}
< \hat{S}(\NN) > = \sum_{p \in Bl}
{\NN}_L(p) C_0
\label{}
\end{equation}

We now discuss the fluctuations. 
For the length operator one can consider 
$\hat{l}_L(c) = \sum_p \hat{l}_L(c_p)$. Here, the sum is over 
black plaquettes which intersect the curve `$c$' and 
$c_p$ refers to the segment of $c$ 
 in the black plaquette $p$. Then, one can easily check that 
\begin{equation}
\Delta \hat{l}_L(c) 
\approx .6 l_P \cos(\theta) \sin(\theta)
\sqrt{N}
\end{equation}
where $N$ is the number of black plaquettes which intersect the curve.
For small $\theta$, 
$\Delta \hat{l}_L(c) \approx .6 l_P \theta \sqrt{N}$. 
This is consistent with the fact that for $\theta =0$ our 
states are eigen states of the length operator. 

In the case of the vector constraint operator 
and the scalar constraint operator 
the calculations are not 
as simple and multiple contributions appear. 
Nonetheless, due to the nature of our state, in the regular domains
a  plaquette is significantly correlated with only a few nearby plaquettes.
The boundaries of the regular domains contribute only a small amount to 
$(\Delta \hat{D}(\vec{N}))^2$  and hence it is easy to verify that 
$(\Delta \hat{D}(\vec{N}))^2$ is proportional to the number of plaquettes
`$N$' in the regular domains. Similar considerations apply to 
$(\Delta \hat{S}(\NN))^2$.  Thus our idea of $\sqrt{N}$ fluctuations
is successfully implemented in the states we have displayed.

One difference (in detail) 
from the calculation of length fluctuations is that 
$\Delta \hat{D}(\vec{N})$ and $\Delta \hat{S}(\NN)$ 
do not vanish when $\theta =0$.
%have terms of order one that are homogeneous in $\theta$. 
For example, an important contribution to 
$(\Delta \hat{D}(\vec{N}))^2$ 
comes from terms of the form 
$ F(p) \cdot L(p) l_P$; we get (to second order in $\theta$) 
\begin{equation}
(\Delta  F(p) \cdot L(p) l_P) \approx 
[ \frac{7}{8} - \frac{7}{16} \theta^2 ] l_P
\label{DeltaD}
\end{equation}
This is when $(p)$ is a black plaquette; for white plaquettes 
we get 
\begin{equation}
(\Delta F(p) \cdot L(p) l_P)\approx 
O(l_P)
\end{equation}
regardless of $\theta$.

%Now we have to investigate whether 
The correspondence between commutator expectation values 
 and Poisson brackets is a more involved calculation and we have not 
investigated this in any detail. Therefore, we restrict ourselves to 
the following remarks.

In the case of the length operators it is easy to see that 
\begin{equation}
< [l_L(c), l_L(c') ] > = 0 . 
\end{equation}
In three spatial dimensions 
there are general reasons to expect that, in quasi-classical 
states, the 
expectation value of the commutator of the area operators 
and its fluctuations 
are small \cite{noncommutativity}; the argument also applies to the 
length operators in the two dimensional case. With regard to the 
calculations for 
$\hat{D}(\vec{N})$ and 
$ \hat{S}(\NN)$, the results of \cite{mvrw1,mvrw2} applied 
to the present context support the correspondence between 
Poisson brackets and commutator expectation values in quasiclassical states.

\subsection*{4.4 Extension to 3+1 dimensions}\label{3d}

There is a natural analog of 
the set of observables that determine our quasi-classicality 
criterion. For the geometry, the area operators and for the connection 
we could consider the induced connection on surfaces with arbitrary 
embedding and measure the connection with the same type of 
``magnetic flux type'' operators (that would have the smearing surface 
as an extra label). This set of operators seems to be 
large enough and would be very close to the 2D case studied here. 
However, we have not done any serious study of its properties. 
Other families may prove to be better. 

Our family of candidate quasi-classical states is tightly tied to a 
two-dimensional 
space. However, the main idea is easily extendible to other dimensions. 
Now we describe it briefly. 

The three-dimensional chess-board geometry inside the regular domains 
is such that black cubes meet only at their vertices. At a vertex ($v_0$) 
two opposite octants are black and the rest are white; one can color the 
whole lattice translating the painted cubes meeting at $v_0$ in the three 
cartesian directions by an even number of steps. 
To each black cell we assign the factor 
$\psi = \cos{\theta} \psi_{2} + \sin{\theta} \psi_{4}$ with 
$\psi_{2n}$ being the spin network state with color $2n$ in the edges of the
black cube. (Other choices with smaller colors are also possible.) 

The factors assigned to the bands in the two dimensional case were found 
using a procedure that can be adapted to the three-dimensional 
case. 
We require that at the boundary of the regular domains the black cells 
share at least three vertices with the other black cells in the domain. 
Then we change the shape of the boundary black cells to have only one 
free vertex. These free vertices are the black vertices needed to construct 
the lattice in the band region and assign a factor of the wave function to 
each band. We use these factors that tie neighboring 
domains with a single spin network of color one per band.

In this way we construct a family of states each of which 
defines a physical lattice. By adjusting the multitude of 
free parameters (density of intersections of the lattice links with 
surfaces that look flat at the microscale 
and $\theta$ as a function of the cells 
of the regular domains) 
we should be able to approximate any given classical data. 
Also, we can restrict to homogeneous states that we would only 
be able to approximate homogeneous classical data.

\section*{5. Discussion}

It is important to clarify that our intent is {\em not}
to provide an alternative quantization to that of loop quantum gravity.
Loop quantum gravity is a theory still under construction 
and thus, a yet incomplete enterprise. Even at the kinematic level,
as we have argued in section 2, the theory is incomplete in that its
most straightforward interpretation does not lead us to the 
classical limit. We view this work as an attempt to remedy this particular
instance of incompleteness by providing a framework to discuss 
quasiclassicality. 
As we have stressed before, this framework is applicable only to the 
calculation of expectation values and fluctuations in quasiclassical
states.
Thus, we do not yet understand the {\em transition} from the
fully quantum regime to the regime in which our framework is proposed to 
apply, namely the semiclassical regime. However, it is clearly necessary to
establish {\em some} framework which defines a satisfactory notion 
of quasiclassicality, before the study of this transition can be 
undertaken and therein, we believe, lies the virtue of this work.

After these preliminary remarks, we discuss
the following issues which arise in the context of our proposal.\\

\noindent {\bf (i)}Superselection sectors: Given a quasiclassical state 
associated with a lattice `$L$', it is clear that no operator of the form 
(\ref{eq:defo}) maps the state out of the space of spin networks based on
$L$. Thus, if operators of the form (\ref{eq:defo}) were the only operators
in loop quantum gravity, we would be faced with the existence of uncountably 
many superselected sectors, one for every choice of lattice. 
However, as stressed in our preliminary remarks above, operators of the form
(\ref{eq:defo}) have been constructed solely to probe the classical limit
in terms of their expectation values and fluctuations in quasiclassical states.
There exist for example, 
in addition to such operators, 
microscopic (Planck scale)
operators in loop quantum gravity which need not be of the form 
(\ref{eq:defo}). Such operators can map quasiclassical states out of the 
putative superselected sectors.

As mentioned earlier in this section, we do not yet understand the relation 
between the fully quantum regime and the semiclassical regime as defined
through our proposal. Hence we do not know the role of these Planck scale
operators in the semiclassical regime. 
Thus, even if there are superselected sectors at the
kinematic level, these sectors may disappear when the dynamical aspects of 
quantum gravity at the Planck scale are incorporated. \\

\noindent {\bf (ii)} Ambiguities in the construction of the operator 
${\hat O}$ corresponding to the classical quantity $O$: On a fixed lattice
`$L$' , there are (infinitely) many microscopically distinct lattice 
approximations to the same continuum quantity. Thus, there are infinitely many,
distinct ways to construct ${\hat O}$ through (\ref{eq:defo}).
It is not clear if we should demand that our state be quasi-classical
with respect to {\em all} possible choices of ${\hat O}$, and if so, 
whether there exist any such states. 
\\

\noindent {\bf (iii)}The algebra of operators of the type $\hat O$:
A qualitatively different ambiguity results from an examination 
of the algebra of operators of the type ${\hat O}$. Let the quasi-classical
state of interest be associated with the lattice $L$.  Consider the operators 
${\hat A}$ and ${\hat B}$ constructed from $A_L$ and $B_L$ through 
(\ref{eq:defo}). For simplicity, assume $[{\hat A},{\hat B}]=0= 
[{\hat A}_L,{\hat B}_L]$.
Then the  operator corresponding to the quantity $AB$ can be 
constructed either as 
\begin{equation}
{\hat {AB}} =\sum_L P_L {\hat A}_L{\hat B}_LP_L
\end{equation}
or as
\begin{equation}
{\hat A}{\hat B} = \sum_L P_L {\hat A}_LP_L{\hat B}_LP_L .
\end{equation}
Since ${\hat{AB}}\neq {\hat A}{\hat B}$, 
there is an ambiguity in the definition of the operator corresponding to 
$AB$. 

There is a special case in which this ambiguity is irrelevant. If the
quasiclassical state based on a lattice is such that $\Delta^{\prime}A$,
$\Delta^{\prime}B$ (see equation (\ref{eq:deltaprime})) are small compared
to $<A>, <B>$ then it can be shown that the 
uncertainty principle implies 
\begin{equation}
{<{\hat {AB}}>\over |<{\hat A}><{\hat B}>|}
= {<{\hat A}{\hat B}>\over |<{\hat A}><{\hat B}>|} \;\;+\epsilon ,
\end{equation}
where 
\begin{equation}
|\epsilon |\leq  {\Delta^{\prime} {\hat A}\over |<\hat{A}>|}
                 \left(1+  ({\Delta^{\prime} {\hat B}\over |<\hat{B}>|})^2
                 \right)^{1\over 2} .  
\end{equation}
Thus, in this special case,
this particular  ambiguity in the definition of the operator corresponding to 
$AB$ is of no consequence.\\

\noindent {\bf (iv)} How small is the microscopic `magnetic flux'?:
Our construction of states with small fluctuations is based on the premise 
that every macroscopic quantity is $N$ times some microscopic quantity with
$N$ very large. Therefore, it is essential that the characteristic 
scale of the microscopic quantity be much smaller than that of the macroscopic 
quantity. In this regard, the `magnetic' flux presents the following 
dilemma. \footnote{Although our arguments involve quantities which are not 
$SU(2)$ gauge invariant, it is easy to see that our conclusions apply to 
any gauge invariant quantities constructed from the magnetic field such as 
$D$ and $S$ of the previous section} 

The classical `magnetic' field is related to the spatial and extrinsic
curvatures through
\begin{equation}
F_{ab}^i = R_{ab}^i + 2\iota D_{[a}K^i_{b]}
+ \iota^2 \epsilon^{i}_{jk}K_a^iK_b^j ,
\label{eq:f=rk}
\end{equation}
where $D_a$ is the operator compatible with the triad and $R_{ab}^i$ is its 
curvature. $K_a^i$ is closely connected with   the extrinsic curvature 
when  all the constraints of general relativity are imposed. If 
 the Immirzi parameter, $\iota$, is of order unity, 
then in any physically reasonable coordinates, it is clear that the classical 
scale for $F^i_{ab}$ is much smaller than an inverse Planck area.
Hence, the magnetic flux through a plaquette of Planck size should be much 
less than unity.  The microscopic flux operator for a Planck size
plaquette `$p$' of the lattice associated  with a quasiclassical state is
$Tr {\hat H}_p\tau^i$. This operator `lives' on the copy of $SU(2)$ 
associated with `$p$' and clearly its fluctuations are of order unity for 
the type of state contemplated in section 4.
This translates to {\em huge} fluctuations of order inverse Planck area 
in the associated microscopic 
magnetic field. Hence the microscopic field fluctuation 
is much larger than the macroscopic scale and our ideas do not apply.
It can be seen that such large fluctuations in the microscopic field, 
for our states, result in large fluctuations in the macroscopic field.
For the macroscopic field averaged over a surface of macroscopic area $A$,
the fluctuations turn out to be of the order ${1\over \sqrt{A}l_P}$ where 
$l_P$ is the Planck length (see equation ({\ref{eq:50})). Thus, the 
macroscopic fluctuations swamp out typical classical values!

Nevertheless, let us see how far we can push our ideas. 
We need to somehow magnify the 
typical macroscopic scale. Notice that this is possible (from (\ref{eq:f=rk}))
if we choose a large value of $\iota$. Then small fluctuations of the extrinsic
curvature magnify to large fluctuations of the magnetic field/flux.
Thus, it is possible to salvage our ideas by appealing to a large $\iota$. 
However, in such a case,it is not clear that the curvature can be identified
with the plaquette holonomy since this identification assumes that the 
plaquette flux is small. Neverthless, if we ignore this objection and choose
$\iota$ to be large and if we still identify $\iota l_{0P}^2$ (in 3d) with the 
Planck area, $l_P^2$, then it is clear that $G_0$ cannot  take the 
value of Newton's constant but must be interpreted as a bare constant. 

Another way to improve matters, say in 3d,
 is to decrease the effective `magnetic field'
by increasing the plaquette size. This is possible if the quasiclassical
state is defined by high spins so that the area of a plaquette is of the 
order of $l^2_{typical} = j_{typical}l_P^2$. Here $j_{typical}$
characterizes the scale of the (high) spins. 
Note that fluctuations in area will then be characterized by
$l^2_{typical}$ rather than $l_P^2$; hence $l_{typical}$ must be much less 
than the macroscopic scale.

For example, in 3d, this idea applied to the (non gauge invariant) 
magnetic flux, $\Phi^i (S)$ (see section 3), through a surface $S$ of area 
$A$ results in the following estimates. Let $N$ be the number of plaquettes
tiling $S$. Then we have $A= Nl_{typical}^2$ and 
$\Delta \Phi^i(S)\sim \sqrt{N}$. Then the fluctuation in the average magnetic 
field $B^i= {\Phi^i(S)\over A}$ is 
\begin{equation}
\Delta B^i := {\Delta\Phi^i(S)\over A} \sim {1\over \sqrt{A l^2_{typical}}}
\label{eq:50}
\end{equation}
instead of $ {1\over \sqrt{A l^2_P}}$.

The emergence of a scale defined by the quasiclassical state between the 
macroscopic scale and the Planck scale can be argued, independently
of our specific ``${1\over\sqrt{N}}$'' inspired constructions.
The area operators (length in 2d) are the fundamental metric dependent 
operators. The uncertainities in the measurement of connection dependent
operators are constrained through the uncertainity principle by the 
size of the fluctuations in the area (length) and the commutator between
the area and the connection dependent operators. The larger the 
permissible uncertainity in the area, the smaller is the achievable 
uncertainity in the connection operators. The scale of 
area fluctuations defines $l_{typical}$ and a characteristic `spin', 
$j_{typical}:= ({l_{typical}\over l_P})^2$. Clearly, $ j_{typical}$  must 
characterize the scale of spins occurring in a spin network decomposition 
of the quasiclassical state. As mentioned earlier, for smoothness of the 
macroscopic geometry, $l_{typical}$ must be much less than the macroscopic 
scale.

A similar picture of the classical limit 
arises in quantum Regge calculus. The relation between the 
Ponzano-Regge-Turaev-Viro partition function and the 
Regge action for three dimensional 
Euclidean spacetimes holds in the large 
$j$ limit \cite{PRmodel,TVmodel}. 
This means that classical smooth spacetimes 
have origin in states whose quantum geometry defines a scale 
$j_{typical} l_P$ which is macroscopically small (to approximate a 
smooth geometry at macroscopic scales) and at the same time is much 
bigger than the Planck scale. 
\\

\noindent {\bf (v)} The possibility of incorporating spatial diffeomorphism
invariance into our proposal: 
Since our constructions do not use any external 
fixed structures, they are covariant with respect to spatial diffeomorphisms.
Hence they ought to generalise to a spatially diffeomorphism 
invariant setting. Such a setting is provided by the Rovelli model 
\cite{carloHK} which combines the Hussain-Kucha{\v r} model \cite{hk} with a
matter reference system. In the context of our constructions, the lattice 
associated with a quasiclassical state for the classical data 
$({\tilde E}^a_{0}, A_{0a})$ can be specified through the choice of a
particular eigenstate of the fermion fields in the Rovelli model.
The fermion fields define  surfaces and the cells of the lattice can be 
located through the intersections of these surfaces. Let us refer to the 
eigenstate of the fermion fields which specifies a lattice $L$  as $|L_F>$.
In the Rovelli model, it is possible to construct classical diffeomorphism
invariant `gravitational' quantities 
by involving the reference matter fields in their 
definition. Our proposal would indicate that  an analysis of the 
classical limit for such diffeomorphism invariant configurations of the
 `gravitational' field and the matter reference system, 
can be done in terms of diffeomorphism invariant operators of the form
\begin{equation}
{\hat O}:=  \sum_L P_{L_F}P_L{\hat O}_L P_LP_{L_F} 
\end{equation}
Here $O_L$ is the lattice approximant of the diffeomorphism invariant
classical quantity $O$ and $P_{L_F}=|L_F><L_F|$ is the projector onto the 
`reference system lattice'. The subsequent considerations of section 3 can be
also be suitably  
generalised to the Rovelli model. The quasiclassical 
state thus constructed will be one for the `gravitational' variables only- 
the matter variables are still very quantum because the 
`matter part of the state', $|L_F>$, 
is an eigenstate of the matter fields.

\section*{6. Conclusions}

In this work we have shown  that there are no quantum states in the 
kinematical Hilbert space of loop quantum gravity which approximate, in 
expectation value, classical holonomies along all possible loops and that,
at best, it may be possible to approximate only a countable number of 
classical holonomies. Since the holonomy variables are the primary 
variables of the loop approach, a new framework to analyse the classical 
limit of kinematic loop quantum gravity is needed which takes into 
account the above result. 

We have proposed a framework in which the choice
of a countable number of loops is made without breaking spatial covariance
by identifying the loops with those which are contained in the graph 
underlying the quasiclassical state itself. Since the graph is required to
be a lattice we are able to import techniques from lattice gauge theory to 
examine various operators of interest (see section 2). 
This part of our work is quite robust.

Next, inspired by the mechanism for low relative fluctuations in 
statistical mechanics, we explicitly constructed candidate quasiclassical
 states in 2 spatial dimensions. Although we could successfully 
implement this mechanism
for low relative fluctuations, the states were not completely satisfactory 
because the fluctuations (as opposed to the relative fluctuations) were
very large. 
More precisely, we were able to construct states which had fluctuations 
of order $\sqrt{N}$ times some naturally occurring microscopic unit, with 
$N$ large. Under the assumption that typical classical values were of 
order $N$ times this unit, these states had ${1\over \sqrt{N}}$ relative 
fluctuations. However, on closer examination we found that this assumption
was unwarranted and that the microscopic unit was not small enough. 
As a result the 
fluctuations swamped out typical classical values.  
Nevertheless, the fluctuations were reduced drastically
in size as compared to the fluctuations at the Planck scale.
For example, at the Planck scale, curvature fluctuations are expected to be
of the order of the inverse Planck area; the mechanism of  low
relative fluctuations reduced the fluctuations in the macroscopic 
curvature by a factor of ${l_P\over \sqrt{A}}$ where $A$ is the macroscopic
area (see {\bf (iv)}, section 5).

Even if our particular explicit construction
of candidate quasiclassical states is irrelevant, it is still true that 
our proposal establishes a connection with lattice gauge theory and 
reinforces  the `weave' based picture of discrete space.
The very fact that we have 
made a connection to lattice gauge theory techniques raises the issue of 
`bareness' of the gravitational coupling 
and the possible existence of several phases and length scales
in our quantum theory. 
In lattice gauge theory, the coupling is renormalized, and phases appear 
due to {\em dynamical} considerations. 
The considerations of the previous section point towards the 
need of considering scenarios for different phases and renormalization of 
coupling constants at the kinematic level. Certainly not much more 
can be inferred in the absence of dynamics, i.e., 
the construction of a projector into the space of physical states 
(in Hamiltonian language, the imposition of the 
diffeomorphism  and, especially, the Hamiltonian constraint).

Since our proposal is new and
unconventional, it is essential to confront our constructions with physically 
reasonable criteria and modify our proposal accordingly. We have attempted to 
do this to some extent in the previous section, but the consequences of our
formalism need to be explored thoroughly before accepting it as a viable 
approach towards an analysis of the classical limit.

Nevertheless, given that a new framework is needed which identifies a
countable set of loops, it seems inevitable that projectors onto this 
set of loops (such as we have defined) play a crucial role. This fact,
along with the need to preserve spatial covariance and the requirement of 
hermiticity of the operator versions of real classical functions, naturally
point towards our specific proposal.

Loop quantum gravity is a very conservative approach to the problem of 
quantum gravity in that it is an attempt to combine the principles of 
quantum mechanics with that of general relativity in accordance with tried
 and tested rules. 
We believe that the real virtue of the loop 
quantum gravity approach is that it captures, in  a clear way, the points of 
tension between quantum mechanics and general relativity and hence 
suggests new ideas beyond the scope of its own framework, 
which may relax this tension. 

In this respect, our work seems to emphasize 
structures {\em intrinsic} to the quantum states as important and hence 
points away from the embedded spin networks of Rovelli and Smolin
\cite{snRS,snBaez} to the 
intrinsically defined spin networks of Penrose \cite{penrose}.
In closing we note that 
the considerations of this work, the qualitative similarity of the resulting 
description of classical space with the quantum statistical 
mechanics description of a classical solid and considerations such as that 
of Jacobson \cite{eqnofstate}, reinforce the idea 
that the dynamics of general relativity 
(and particularly the Hamiltonian constraint) may arise as a 
coarse grained/statistical description of fundamental degrees of freedom at 
the Planck scale. \\

\vspace{5mm}

{\bf Acknowledgements}: We are indebted to the anonymous referee for 
invaluable comments and suggestions. JAZ was partially supported by 
CONACyT-990443.

\section*{Appendix}
\subsection*{A1}
Let the space of loops with base point $x_0$ be ${\cal L}_{x_0}$.
Denote the trivial loop by $e$.
As in \cite{ai}, consider the free vector space ${\cal FL}_{x_0}$ 
generated by loops in ${\cal L}_{x_0}$. On ${\cal FL}_{x_0}$,
define the product law
\begin{equation}
(\sum_{i=1}^n a_i \alpha_i)(\sum_{j=1}^m b_j \beta_j)
:= (\sum_{i=1}^n\sum_{j=1}^m a_ib_j \alpha_i\circ \beta_j)      
\end{equation}
and the involution
\begin{equation}
(\sum_{i=1}^n a_i \alpha_i)^{\dagger}
:=\sum_{i=1}^n a_i^* \alpha_i^{-1} .
\end{equation}
Here,
 $a_i, b_j$ are complex numbers and $\alpha_i, \beta_j \in {\cal L}_{x_0}$.

Fix a connection $A_{0a}^i$ and
extend the definition of holonomy trace 
to ${\cal FL}_{x_0}$ by linearity so that
\begin{equation}
T^0_{\sum_{i=1}^n a_i\alpha_i}(A_0)
=\sum_{i=1}^n a_iT^0_{\alpha_i}(A_0) .
\end{equation}
Next, define 
\begin{equation}
I_{A_0}:= \{ \sum_{i=1}^n a_i\alpha_i \in {\cal FL}_{x_0}|
     \sum_{i=1}^n a_iT^0_{\alpha_i\circ \beta}(A_0)= 0 \;\rm{for\; every}\;
   \beta \in {\cal L}_{x_0} \} .
\end{equation}
It can be checked that $I_{A_0}$ is a two sided ideal in ${\cal FL}_{x_0}$.

Note that, since $T^0$ is an $SU(2)$ trace, under involution 
\begin{equation}
T^0_{(\sum_{i=1}^n a_i\alpha_i)^{\dagger}}(A_0)=
\sum_{i=1}^n a_i^*T^0_{\alpha_i}(A_0) 
=(\sum_{i=1}^n a_iT^0_{\alpha_i}(A_0))^* .
\label{eq:invol}
\end{equation}
We choose $A_0$ such that there exists some loop $\tau\in {\cal L}_{x_0}$ 
for which
\begin{equation}
|T^0_{\tau}(A_0)| \neq 1. 
\label{eq:r1}
\end{equation}
Define the complex numbers 
$l_1(\tau ), l_2 (\tau )$ as
\begin{equation}
l_1(\tau ):= T^0_{\tau}(A_0) + i (1-(T^0_{\tau}(A_0))^2)^{1\over 2},
\end{equation}
\begin{equation}
l_2 (\tau ):= T^0_{\tau}(A_0) - i (1-(T^0_{\tau}(A_0))^2)^{1\over 2}
\end{equation}
and $\rho_1(\tau ), \rho_2(\tau ) \in {\cal FL}_{x_0}$ as
\begin{equation}
\rho_1(\tau ) := (l_1(\tau )- l_2(\tau ))^{-1} (\tau - l_2 e),
\end{equation}
\begin{equation}
\rho_2 (\tau ):= (l_2(\tau )- l_1(\tau ))^{-1} (\tau - l_1 e).
\end{equation}
It can be checked that mod $I_{A_0}$, 
\begin{equation}
\rho_i(\tau ) \rho_j(\tau ) =\delta_{ij} \rho_i(\tau ), 
\;\;\;\;\;\; \rho_1(\tau ) + \rho_2(\tau )= e
\label{eq:rhoproj}
\end{equation}
and that for any $\alpha\in {\cal L}_{x_0}$
\begin{eqnarray}
T^0_{\rho_1(\tau^{-1}) \alpha }(A_0) &= &T^0_{\rho_2(\tau )\alpha}(A_0) .
\label{eq:rho-1} \\
T^0_{(\rho_1(\tau ))^{\dagger} \alpha }(A_0) &= 
&T^0_{\rho_1(\tau )\alpha}(A_0) .
\label{eq:rho+} 
\end{eqnarray}
We shall further restrict attention to $A_{0a}^i$ such that 
there exists some $a \in {\cal FL}_{x_0}$ for which 
\begin{equation}
C:=T^0_{\rho_1(\tau  ) a \rho_2 (\tau )a^{\dagger}}(A_0) \neq 0.
\label{eq:r2}
\end{equation}
Using the algebraic properties of the $T^0$ variables and (\ref{eq:invol}),
(\ref{eq:rhoproj}), (\ref{eq:rho-1}) and (\ref{eq:rho+})
 it can be verified that 
\begin{equation}
U_{\alpha}(A_0):= 
\left( \begin{array}{ll}
2T^0_{\alpha \rho_1(\tau )}(A_0) & 
{2T^0_{\rho_1(\tau )\alpha \rho_2(\tau )
                              a^{\dagger}}(A_0)\over \sqrt{2C}}\\
{2T^0_{\rho_2(\tau )\alpha \rho_1(\tau )a}(A_0)\over \sqrt{2C}}
                          & 2T^0_{\alpha \rho_2(\tau )}(A_0)
\end{array}
\right)
\label{eq:Udef}
\end{equation}
is an $SU(2)$ matrix such that 
$U_{\alpha}(A_0)U_{\beta}(A_0) =U_{\alpha\circ \beta}(A_0)$ with 
${1\over 2} Tr U_{\alpha}(A_0)= T^0_{\alpha}(A_0)$.
Details of this construction maybe found in \cite{giles}.

We note that the proof of the above properties of $U_{\alpha}(A_0)$ 
depend solely on the {\em algebraic} properties of the $T^0$ (and their 
extensions to ${\cal FL}_{x_0}$) and  the property (\ref{eq:invol}) 
of the $T^0$ under involution; and is independent of the particular
 connection $A_0$.
\footnote{
Provided, of course, that the various expressions in
(\ref{eq:Udef}) are well defined. That they are indeed well-defined 
is guaranteed by the requirements (\ref{eq:r1}) and (\ref{eq:r2}).}
These algebraic properties are shared by the  ${\hat T}^0$
operators and the property (\ref{eq:invol}) translates to adjointness 
properties of the ${\hat T}^0$ operators. 
Moreover, since these operators form a commutative algebra
it can be verified that substituting ${\hat T}^0$ for all occurrences of 
$T^0(A_0)$ 
in the above
construction, yields an $SU(2)$ valued operator ${\hat U}_{\alpha}$
such that 
${\hat U}_{\alpha}{\hat U}_{\beta} ={
\hat U}_{\alpha\circ \beta}$ and 
${1\over 2} Tr {\hat U}_{\alpha}= {\hat T}^0_{\alpha}$.
\footnote{
The counterpart of the restrictions (\ref{eq:r1}) and (\ref{eq:r2})
is the fact that some of the operators 
encountered ( namely, $({\hat l}_1(\tau )- {\hat l}_2(\tau ))^{-1}$
and ${\hat C}^{-{1\over 2}}$) are unbounded.
We assume that mathematical subtleties related to domain issues of 
unbounded operators can be taken care of in a more careful treatment.
}

Now we can substitute ${\hat H}$ by ${\hat U}$ in the arguments of section 2
and obtain (\ref{eq:t0=0}), this time, in a gauge invariant context with the 
(weak) restrictions (\ref{eq:r1}) and (\ref{eq:r2}) on the classical 
connection $A_{0a}^i$.

%%%%%%%%%%
  \nocite{*}                   %this uses *everything* in the .bib file
          \bibliography{biblio}        %or whatever your .bib file is
          \bibliographystyle{utphys}   %if you use utphys.bst
%%%%%%%%%%
\end{document}